# An artificial spiking synapse made of molecules and nanoparticles

F. Alibart,[1] S. Pleutin,[1] D. Guerin,[1] C. Gamrat,[2] D. Vuillaume.[1*]

1. Molecular Nanostructures and Devices group, Institute for Electronics Microelectronics and Nanotechnology, CNRS, University of Lille, BP60069, avenue Poincaré, F-59652cedex, Villeneuve d'Ascq, France.

2. CEA, LIST/LCE (Advanced Computer technologies and Architectures), Bat. 528, F-91191, Gif-sur-Yvette, France.

Molecule-based devices are envisioned to complement silicon devices by providing new functions or already existing functions at a simpler process level and at a lower cost by virtue of their self-organization capabilities, moreover, they are not bound to von Neuman architecture and this may open the way to other architectural paradigms (1). Here we demonstrate a device made of conjugated molecules and metal nanoparticles (NPs) which behaves as a spiking synapse suitable for integration in neural network architectures. We demonstrate that this device exhibits the main behavior of a biological synapse. The device (Fig. 1A) consists of a bottom-gate, bottom source-drain contact organic transistor configuration. The gold NPs (5 nm in diameter) were immobilized into the source-drain channel using surface chemistry (self-assembled monolayers) and they were subsequently covered by a thin film (25-35 nm thick) of pentacene (see supporting online material). This device gathers the behavior of a transistor and a memory (2) and it is referenced to as NOMFET (Nanoparticle Organic Memory Field Effect Transistor).

The most important feature of a synapse is its ability to transmit in a given way an action potentials (APs) from one pre-synapse neuron N1, to a post-synapse neuron N2. When a sequence of APs is send by N1 to N2, the synaptic behavior determines the way the information is treated. The synapse transforms a spike arriving from the presynaptic neurone into a chemical discharge of neurotransmitters that will be detected by the post-synaptic neurone and transformed into a new spike. Markram and Tsodyks (3, 4) have proposed a phenomenological model to describe the synapse behavior. The synapse possesses a finite amount of resources: the chemical neurotransmitters. Each spike activates a fraction of these resources and the amplitude of the transmitted spike is proportional to this fraction. The fraction of neurotransmitters spend to transmit the information is then recovered with a characteristic time $\tau_{rec}$ that is typically in the range of the second. The response of a synapse to a train of pulses with variable frequency can be calculated by an iterative model (5) which describes the biological synapse behavior reasonably well (Fig. 1B). The main feature of a biological synapse is to present a dependence of the amplitude of the output spike with the frequency of the input spike. It also depends on the history of the synapse which determines the amount of available neurotransmitter at a given time. Such a typical behavior is shown in Fig. 1B: at high(low) frequency, the period of the input signal is lower(larger) than $\tau_{rec}$ and the output signal decreases(increases) at each successive pulse generating a depressing(facilitating) behavior (Fig. 1B).

We used the NOMFET as a "pseudo two-terminal device". The gate receives the same input voltage (a train of pulse at frequency 1/T, amplitude V, and pulse width W) as the source electrode. The output is the drain current (Fig. 1A). We measured the response of the NOMFET to sequences of pulses with different periods, T (Fig. 1C). During such experiments, the NPs are alternatively charged during the pulse duration and discharged between pulses (2). The value of the current at a certain time depends on the full history of the device that determines the amount of charges presents in the NPs. To illustrate this point let us consider the system at the beginning of a particular sequence with period T (Fig. 1C), where the NPs contain some charges. If $T<<\tau_d$ ($\tau_d$ is the NP discharge time constant of about 20 s here), more and more holes are trapped in NPs and the NOMFET presents a depressing behavior. Then, for a larger period T (Fig. 1C), NPs have enough time to be discharged between pulses and the sequence presents a facilitating behavior. This feature exactly reproduces the behavior of a biological synapse. The holes trapped in the NPs play the role of the neurotransmitters and the output signal, $I_D$, is a decreasing function of the number of holes stored in the NPs (2). At each spike, a certain amount of holes are trapped in the NPs. Between pulses the system relaxes: the holes escape with a characteristic time $\tau_d$. This behavior persists when shrinking the NOMFET to a source-drain channel of 1 μm (Fig. 1D). In that case, we used a constant dc bias on the common gate (due to channel reduction, the lateral field is sufficient to charge and discharge the NP without the need to applied the pulse on the gate electrode). This features would simplify the architecture design of neural networks using NOMFETs since a separate gate is not required for each NOMFET. Note that the depressing/facilitating behaviors are now inverted (with respect of the frequency of the pulses) and the time constant decreased- see supporting online material. Finally, such a synaptic behavior was not observed for the reference pentacene OFET (no NPs) - Fig. S2.

These results open the way to rate coding utilization (6) of the NOMFET in perceptron and Hopfield networks (7). We can also envision the NOMFET as a building block of neuroelectronics for interfacing neurons or neuronal logic devices made from patterned neuronal cultures with solid-state devices and circuits (8, 9).

* To whom correspondence should be addressed. E-mail: dominique.vuillaume@iemn.univ-lille1.fr

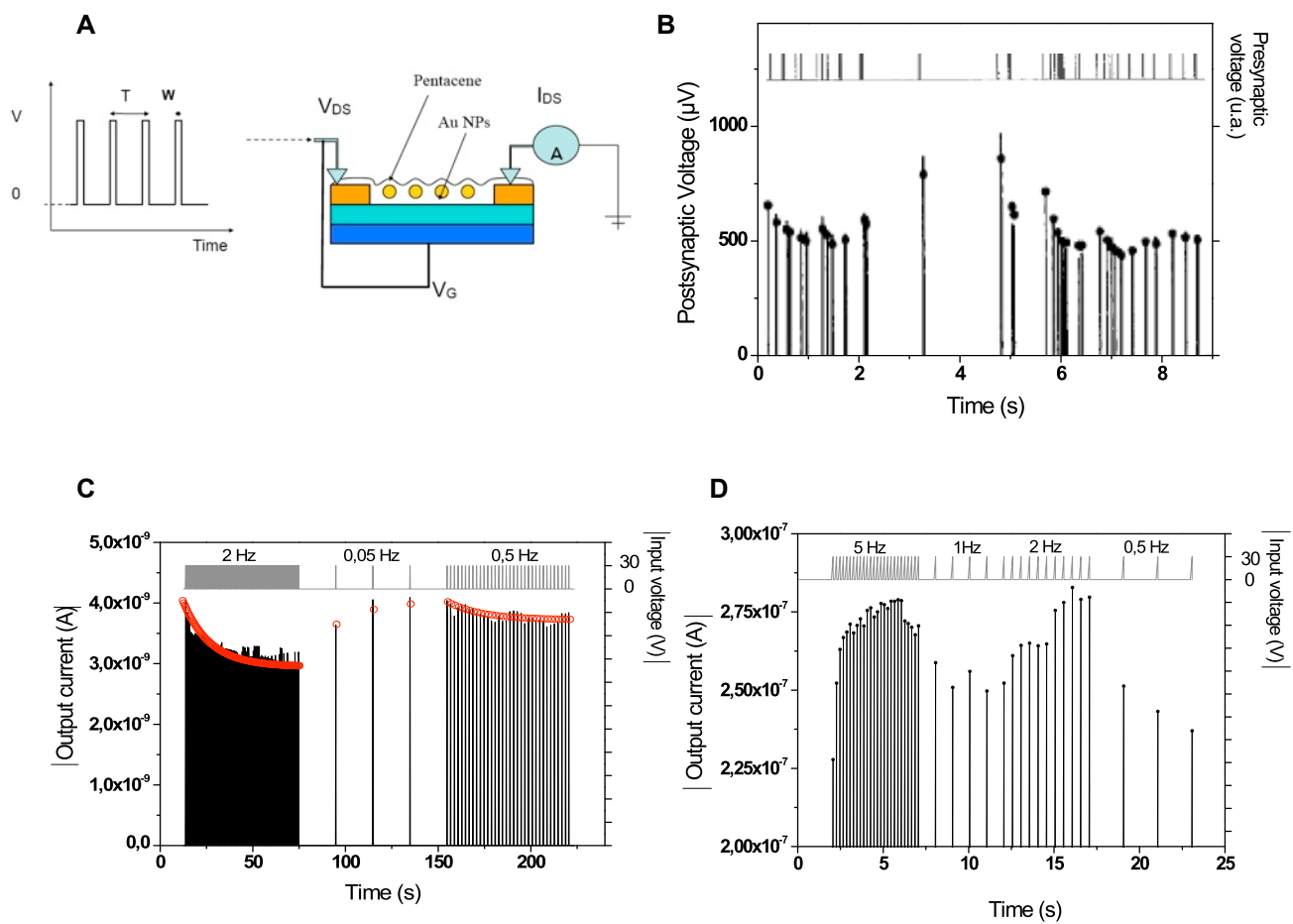

**Fig. 1. (A)** Schematic view of the NOMFET and measurement configuration. **(B)** Comparison between the measured and simulated behaviors of a synapse in layer 2/3 of rat primary visual cortex (from Ref. (5)). **(C)** Typical responses of a large (12 µm) NOMFET excited by a voltage pule (-30 V, 100 ms) at several frequencies. **(D)** Typical response of a 1 µm NOMFET excited with the same voltage pulse on the source electrode, while the gate is at a fixed bias of -20V.

*Supporting online material for*

**An artificial spiking synapse made of molecules and nanoparticles**

F. Alibart,[1] S. Pleutin,[1] D. Guerin,[1] C. Gamrat,[2] D. Vuillaume.[1*]

1. Molecular Nanostructures and Devices group, Institute for Electronics Microelectronics and Nanotechnology, CNRS, University of Lille, BP60069, avenue Poincaré, F-59652cedex, Villeneuve d'Ascq, France.

2. CEA, LIST/LCE (Advanced Computer technologies and Architectures), Bat. 528, F-91191, Gif-sur-Yvette, France.

* To whom the correspondence should be addressed. E-mail: dominique.vuillaume@iemn.univ-lille1.fr

**Materials and methods**

The NOMFET are processed using a bottom-gate electrode configuration. We used highly-doped ($\sim 10^{-3}$ Ω.cm) p-type silicon covered with a thermally grown 200 nm thick silicon dioxide. Before use, these wafers were cleaned by a piranha solution ($H_2SO_4$ /$H_2O_2$, 2/1 v/v) for 15 minutes and then ultraviolet irradiated in ozone atmosphere (ozonolysis) for 30 minutes (**Caution:** *piranha solution is highly exothermic and reacts violently with organics*). Networks of square-shaped gold electrodes (113 µm sides, inter electrode gaps of 12 µm) were deposited on the substrate by vacuum evaporation of titanium/gold (20/200 nm) through a shadow mask. Networks of electrodes with a smaller gap (1 µm) were fabricated by usual photo-lithography.

Then, the $SiO_2$ (gate dielectric) was functionalized by self-assembled monolayer (SAM) of a thiol-ended molecule. The SAM was prepared by a silanization reaction in gas phase. The oxidized silicon with gold electrodes wafer was placed overnight in the presence of vapors of freshly distilled mercaptopropyltrimethoxysilane (MPTS) in a laboratory glassware at 0.2 Torr. This freshly prepared substrate was immersed in a gold nanoparticles (NPs) solution. We used a solution of 4-5 nm (in diameter) dodecanethiol functionalized gold nanoparticles (2 % in toluene) supplied by Aldrich. This starting solution is diluted 100 times in toluene. As expected, thiol capped Au NPs readily react with thiol-terminated SAM by ligand exchange forming a covalent bond with the surface. We present in figure S1 a scanning electron microscopy (SEM) image of the inter-electrode gap. We obtained a rather uniform distribution of NPs (no NP aggregation) with a density of about 6.5x10$^{11}$ NP/cm$^2$. Finally, a 35 nm thick pentacene film was evaporated at a rate of 0.1 Å/s. A reference device of pentacene without NPs was also realized in the same run of deposition to evidence the effect of NPs on the electrical properties.

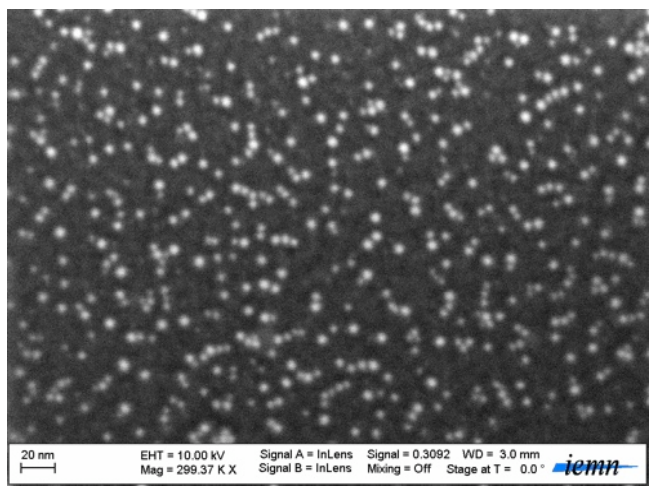

Fig. S1: SEM image of the Inter-electrode region covered with Au NPs attached on the thiolated SAM. The NPs density (about 6.5x10$^{11}$ cm$^{-2}$) is homogeneous on the entire sample.

The NOMFET electrical characteristics were measured with an Agilent 4155C semiconductor parameter analyzer, the input pulses were delivered by a pulse generator (Tabor 5061). The electrodes of the NOMFET were contacted with a micro-manipulator probe station (Suss Microtec PM-5) placed inside a glove box (MBRAUN) with a strictly controlled nitrogen ambient (less than 1 ppm of water vapor and oxygen). The organic semiconductor (pentacene) being a hole conducting material (p-type semiconductor), the NOMFET is active for negative voltages only. We measured the output drain current $I_D$ versus time with a train of negative voltage pulses applied on the source.    We used a pulse amplitude of - 30 V, a pulse width of 100 ms, the pulse frequency was in the range 0.03 to 5 Hz (according to the typical charge/discharge time constants of the NPs measured elsewhere (2)).  For the large NOMFET (12 µm) the source and gate are short-circuited so as to applied the train of pulses on the gate electrode also. This mandatory for an efficient charging and discharging of the NPs at the SiO$_2$/pentacene interface through the applied gate oxide field. For the short NOMFET (1 µm and below), the lateral electric field between source and drain is high enough to induce the charging and discharging of the NPs, and therefore a constant voltage (-20V) was applied to the gate electrode to turn the device on.

**Supporting text.**

In the 1 µm NOMFET, the depressing/facilitating behaviors are now inverted (with respect of the frequency of the pulses) - Fig. 1D. This feature comes from the fact that a constant negative bias (- 20 V) is applied on the gate which corresponds between two pulses to a charging configuration for the NPs. When a larger negative pulse (here - 30 V) is applied on the source, the source-gate field tends to discharge the NPs, leading to the observed increase in the drain current. The typical time constant also decreased (compare time scale between Fig. 1C and 1D), and are now more closer than the one in a biological synapse (Fig. 1B). This is mainly due to the reduction of the channel size and a decease in the channel resistance (higher drain current). This feature reduces the time constant R.C, where the capacitance is mainly the capacitance of the gold NPs and R the resistance of the organic channel.

**Supporting figure.**

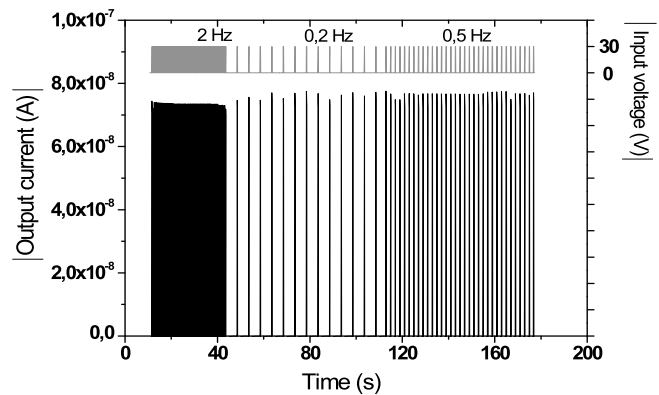

Fig. S2. Response of a reference device (no NP incorporated in the pentacene layer) to a train of pulses. No synaptic behavior is observed.

1.  C. Novembre, D. Guérin, K. Lmimouni, C. Gamrat, D. Vuillaume, *Appl. Phys. Lett.* **92**, 103314 (2008).